\documentclass[doublecol]{epl2}

\title{The role of tap duration for the steady state
density of vibrated granular media}

\author{Joshua A. Dijksman\inst{1} \and Martin van Hecke\inst{1}}
\shortauthor{Joshua A. Dijksman and Martin van Hecke}

\institute{\inst{1} Kamerlingh Onnes Lab, Universiteit Leiden,
Postbus 9504, 2300 RA Leiden, The Netherlands}

\pacs{45.70.-n}{}

\pacs{45.70.Cc}{}

\pacs{81.20.Ev}{}

\usepackage{graphicx}
\usepackage{dcolumn}

\usepackage[]{amsmath,amssymb,amsfonts}

\abstract{ We revisit the problem of compaction of a column of
granular matter exposed to discrete taps. We accurately control
the vertical motion of the column, which allows us to vary the
duration $T$ and the amplitude $A$ of single-cycle sinusoidal taps
independently. We find that the density of the material at the
reversible branch depends both on $A$ and $T$. By comparing the
densities on the reversible branches obtained for a range of
values of $T$, we find that we can collapse all data when plotted
as function of $A/T$, which scales similar to both the liftoff
velocity and the time of flight of the packing. We further show
that switching between states obtained for different $A$ and $T$,
but chosen such that their densities on the reversible branches
match, does not lead to appreciable hysteresis. We conclude that
the appropriate control parameter for sinusoidal tapping is not
the peak acceleration $\Gamma \sim A/T^2$, as is usually assumed,
but rather $\Gamma T \sim
 A/T$.}

\begin{document}

\maketitle
\date{\today}

Everyday experience tells us that granular materials compact when
vibrated --- think of tapping a tin of coffee powder in a can.
Early experiments in the 1950s indicated that tapping can induce
both dilation and compaction of a packing of granular materials,
depending on the details of the tap ~\cite{1957_nature_macrae}.
More than a decade ago, a series of experiments probed the
compaction of dry granular materials (glass beads) in a narrow
tall tube which was tapped
vertically~\cite{1995_pre_knight,1997_powtech_nowak,1998_physd_bennaim}.
Starting from a loose packing, the density was observed to exhibit
slow, logarithmic growth as a function of the number of applied
taps. Memory effects, in which the density evolution depended on
the tapping history, were also found~\cite{2000_prl_josserand}.

Eventually however, in all experiments a state was reached where
the packing density depends only on the tapping strength and not
on the history~\cite{2005_naturemat_richard}. From such a state
the so-called reversible branch could then be obtained repeatedly
and reversibly by slowly ramping up and down the tapping
intensity. The transient phenomena and memory effects all occur
along the so called irreversible branch, along which the system
would evolve before reaching the reversible branch.

In later studies of compaction in much wider containers,
convection was found to be
important~\cite{2003_prl_philippe,2003_pre_richard,2002_epl_philippe}.
In these experiments the temporal evolution of the density on the
irreversible branch was found to be
different~\cite{2005_naturemat_richard,2003_prl_philippe,2003_pre_richard,2002_epl_philippe},
but again the same reversible branch was found.

Here we address the following question: What is the appropriate
control parameter that characterizes the taps? A widely used
characterization of taps is the ratio of their peak acceleration
and the gravitational acceleration, $\Gamma$. Certainly the peak
acceleration is important in that it allows to distinguish between
tap strengths where no liftoff of the packing occurs, for $\Gamma
< \Gamma^*\approx 1$, and taps where that does happen
\cite{2003_prl_philippe}. Recent numerical work on the
irreversible branch dynamics of compaction has suggested that the
dimensionless acceleration parameter is not appropriate for
rescaling the data \cite{2008_epl_ludewig}. Moreover, recent
numerics~\cite{2008_powtech_an} of compaction under sinusoidal
driving vibrations indicate that the vibration frequency also
influences the reversible branch, and similarly, supporting
evidence for the role of tap duration can already be found in the
observations of Macrae {\em et al.} ~\cite{1957_nature_macrae}.

Here we address the question of the appropriate control parameter
experimentally, by studying the packing density on the reversible
branch in experiments in which we expose granular packings to
single cycle sinusoidal discrete taps with $\Gamma > 1$, where we
control and vary both the tap amplitude $A$ and its duration $T$
(Fig.~\ref{fig:setup}). For sinusoidal taps as used here, $\Gamma$
can be given in terms of $A$ and $T$: $\Gamma=A \omega^2/g$, with
$\omega=2\pi/T$ the radial frequency, and $g$ the gravitational
acceleration. Since we have precise control over the acceleration
signal, we can vary both $\Gamma$ and $\omega$ independently (see
Fig.~\ref{fig:setup}b).

For given $T$, we obtain similar reversible branches $\phi_{\rm
rev}(\Gamma,T)$ as were observed before for fixed $T$, both for a
bidisperse glass bead mixture and a bronze powder, and the
reversible densities $\phi_{\rm rev}(\Gamma,T)$ depend both on $T$
and $\Gamma$. We find that all data can be collapsed in good
approximation by plotting the packing fraction as function of
$\Gamma T$, which is close to the liftoff velocity, or similarly,
the time of flight of the granular packing. In addition we probe
what happens when we alternate the taps between pairs of different
$\Gamma$ and $T$ for which $\phi_{\rm rev}(\Gamma,T)$ are equal,
and we find no appreciable hysteresis.

The evidence that we will detail below will show that $\Gamma$ is
not the appropriate control parameter for compaction. Our results
suggest that the density of the reversible branch is controlled by
the product of $\Gamma$ and large $T$. Hence, by maximizing
$\Gamma~T$ instead of $\Gamma$, a wider range of agitation
strengths can be reached.

\section{\label{sec:setup}Experimental Methods}
\begin{figure}[t]
    \includegraphics[width=8cm]{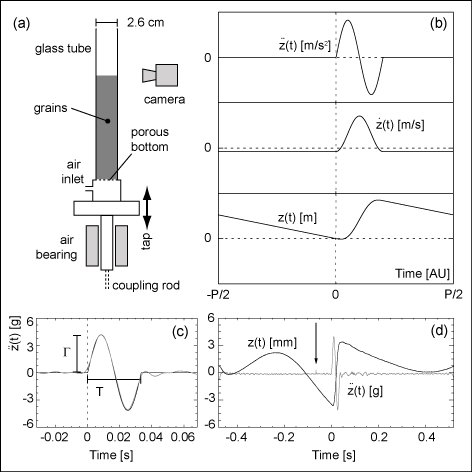}
    \caption{\label{fig:setup} {\bf} (a) Sketch of the experimental setup. (b)
    Sketch of the waveforms of $\ddot{z}(t)$, $\dot{z}(t)$ and
    $z(t)$, illustrating the offset and linear slopes present in
    $z(t)$, stemming from continuity requirements. (c)
    Comparison between desired waveform (grey) and actual waveform
    (black) of the vertical acceleration $\ddot{z}(t)$. (d)
    A typical measured position (black) and acceleration (grey) signal.
    The arrow indicates the phase at which the
    packing fraction is measured.}
\end{figure}
{\em Setup.} We study granular compaction in a glass tube
(diameter 26 mm, height 20 cm) filled with grains (typical filling
height 10 cm). The glass tube is shaken vertically with a shaker
(VG100, VTS systems), driven by a commercial audio amplifier
(Crown CE1000). The vertical motion of the tube is guided by an
air bearing ($\varnothing$ 1", New Way) which is levelled.
Levelling eliminates heaping, the unwanted tilting of the surface
of granular media subjected to vibrations ~\cite{1995_prl_pak}. An
0.5 meter long and flexible aluminum rod couples the tube and the
shaker (see Fig.~\ref{fig:setup}a). This rod eliminates the
necessity of excessively accurate alignment of the axes of the air
bearing and the shaker.

The vertical acceleration of the tube is measured with an accuracy
of $10^{-3}$g, by a combination of two accelerometers (Dytran
E3120AK and a modified ADXL320EB). For consistency checks, we
occasionally monitor the position $z(t)$ of the glass tube with an
inductive position sensor (Messotron WLH50, $10^{-5}$m resolution)
--- the acceleration measurements are far more sensitive and these
will be used in the feedback scheme described below.


{\em Tap ---} In order to create a tap where both the duration and
strength can be controlled independently, we need to determine the
waveform $z(t)$. We denote the duration of the tap by
$T=2\pi/\omega$, the period of the tap by $P$ (we will fix it
later at 1 second), and have the tap start at $t=0$. We demand
than that $z(t)$, $\dot{z}(t)$ and $\ddot{z}(t)$ are continuous,
that $\int dt \dot{z}(t)=0$ and $\int dt \ddot{z}(t)=0$, and that
$z(0)=0$. These requirements severely constrain the tap --- for
example, taking for $z(t)$ a single sine cycle followed by a
period where $z=0$ makes $\ddot{z}(t)$ discontinuous. Starting
from a single sine cycle for $\ddot{z}(t)$, obtaining $\dot{z}(t)$
and ${z}(t)$ by integration, and fixing the integration constants
so that the continuity conditions are full filled, we find the
waveform summarized in Table~\ref{wavetable} (see
Fig.~\ref{fig:setup}b):


\begin{table}[ht]
    \begin{center}
        \begin{tabular}[c]{|l|l|l|}
        \hline
        & $0 \leq t \leq T$ &$T \leq t  \leq P$ \\
        \hline
        $\ddot{z}~~~~~~~~$ & $A \omega^2 ~sin (\omega t)$ & $0$\\
        $\dot{z}$ & $-A \omega ~cos (\omega t) + b_1$ & $-b_2$\\
        ${z}$ & $-A ~sin (\omega t) + b_1 t$ & $-b_2 (t-P)$\\
        \hline
        \end{tabular}
        \caption{Waveform of the tap. Here the integration constants equal
        $b_1=A \omega (P-T)/P$ and $b_2=a \omega~ T/P$.} \label{wavetable}
    \end{center}
\end{table}

In order to create such a tap, a feedback algorithm is used to
adapt the Fourier components of the signal fed into the audio
amplifier such that the measured acceleration signal converges to
the desired waveform. This procedure is carried out once, before
the start of the experiment, and the shape of the output wave is
not changed during the actual experiment; it is only multiplied by
a scale factor to set its overall amplitude. We have checked that
the system is sufficiently linear so that re-calibration of the
waveform is not necessary when $\Gamma$ is varied.

Taps are applied once every second, so $P = 1$; this allows the
packing to come completely to rest before the start of a new tap.
In Fig.~\ref{fig:setup}c, the desired waveform for $\ddot{z}(t)$
is compared to a typical measured acceleration signal
--- the waveform produced by the feedback scheme
is in good agreement with the desired tap. In
Fig.~\ref{fig:setup}d, a measured waveform for $z(t)$ and
accompanying $\ddot{z}(t)$ are shown. The slow downward motion of
the tube anticipated in Table~\ref{wavetable} is clearly visible.
This downward motion is not at constant speed due to low-frequency
limitations in the electronics that drive the shaker. The
accelerations associated with this non-constant speed are however
always below $g$, so they do not influence the packing fraction
measurably.

\begin{figure}[t]
    \includegraphics[width=8cm]{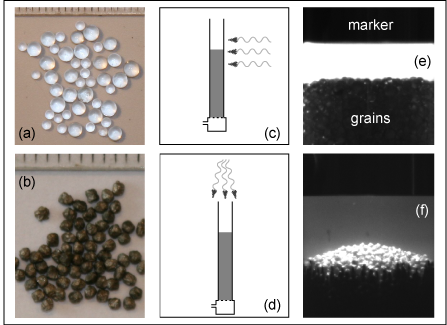}
    \caption{\label{fig:bronzeglass} (Color online) The granular
    material used in the experiment (scale = 1mm): (a) glass
    beads, (b) bronze powder. Illumination direction: (c) for glass, from the
    back, (d) for bronze, from the top. Typical camera image (e)
    glass, (f) bronze. For bronze powder two cameras are used, image is
    typical for both.}
\end{figure}

{\em Parameter Range and Grain Dynamics.} As was established in
\cite{2003_prl_philippe}, the dynamics of compaction dramatically
slows down when the tapping peak acceleration $\Gamma$ falls below
a critical value $\Gamma^*$ which is typically slightly above 1.
This transition is accompanied by a change from a regime where the
grains experience lift off for $\Gamma > \Gamma^*$ to a regime
where they do not. In our experiments, the values of $\Gamma$
range from $1$ to $15$, and by high speed imaging we have
established a clear lift off and expansion of the whole column of
grains for $\Gamma \gtrsim 2$ --- the precise value of $\Gamma^*$
in our experiments is likely somewhat smaller. The vast majority
of our data is therefore taken in the regime where the grains
loose contact with the bottom plate and their neighbors.

{\em Packing Density and Material Used.} The packing density is
determined from the height of the granular column, and this height
is measured with a camera that observes the packing from the side,
as in Refs.~\cite{2005_pre_schroter,2008_prl_divoux}. The camera
is triggered for strobed image acquisition, such that the camera
takes a picture just before the start of the tap: the trigger
moment is indicated by the arrow in Fig.~\ref{fig:setup}d. A
ring-shaped dark marker, whose height from the bottom of the tube
is known (typically 13 cm), is attached to the tube, slightly
above the maximum height of the bead pack. Depending on the grains
used we use lighting from behind (for glass beads), or from the
top (for bronze powder) --- see Fig.~\ref{fig:bronzeglass}.
Determining the height of the packing in both cases amounts to
counting pixels with intensity above a certain threshold value.
The threshold value is determined from the histogram of a typical
image. We verified that the imaging method applied gives a linear
relationship between the amount of grains in the tube and the
number of pixels in the gap.\\

We employ two types of granular matter: a bidisperse mixture of
glass beads (Pneumix, 1.6 and 2.3 mm in 1:1 volume mixture,
Fig.~\ref{fig:bronzeglass}a), or monodisperse bronze powder
(Acupowder, grade '12HP', $\sim$ 1mm diameter,
Fig.~\ref{fig:bronzeglass}b). Due to their different opacity,
optimal lighting is different for these materials
(Fig.~\ref{fig:bronzeglass}c-d). One more difference is that glass
bead packings generally have a flat surface, while the bronze
beads grow a small radially symmetric heap at their surface (this
heap is likely due to convection). One important difference is
that the glass beads tend to be more sensitive to triboelectric
effects, which result in overall, run-to-run variations of the
absolute density (the trends in the variation of $\phi_{\rm rev}$
with $\Gamma$ are not affected by such variations). We observe
that such variations are absent for the bronze powder.
Importantly, our essential findings (control parameter $\sim
\Gamma T$ and no hysteresis) are seen with both materials.

Our measurement method yields a high resolution; the noise in the
determination of the height in the packing due to camera pixel
noise alone is 30$\mu$m, this translates into a variation of
$\phi$ of $\sim 0.03\%$. However, this method gives a relatively
poor accuracy in the absolute value of $\phi$. This is due to the
cumulative effect of the errors in the determination of the height
of the marker, the inner diameter of the tube, the density of the
beads, image calibration and the thresholding. The packing
fractions stated in this paper therefore have an estimated
systematic error of 1.2$\%$.

{\em Experimental Protocol.} The bottom of the tube is perforated
and connected to a flow box so dry air can be pumped through the
bead pack. We always start an experiment by fluidizing the packing
with several pulses of dried compressed air. This procedure
creates an initial packing density of order $0.60 \pm 0.01$. The
number of flow pulses is generally not the same for each
experiment; several pulses are applied initially to ensure a
proper erasure of memory effects~\cite{1995_pre_knight,
2008_prl_divoux}, after which pulses are applied until a packing
with a flat surface is obtained. Airflow is turned off during all
compaction experiments, which simply consist of
observing how $\phi$ evolves while taps are applied.\\

\section{\label{sec:pheno}Transients \& Steady State}

\begin{figure}
    \includegraphics[width=8cm]{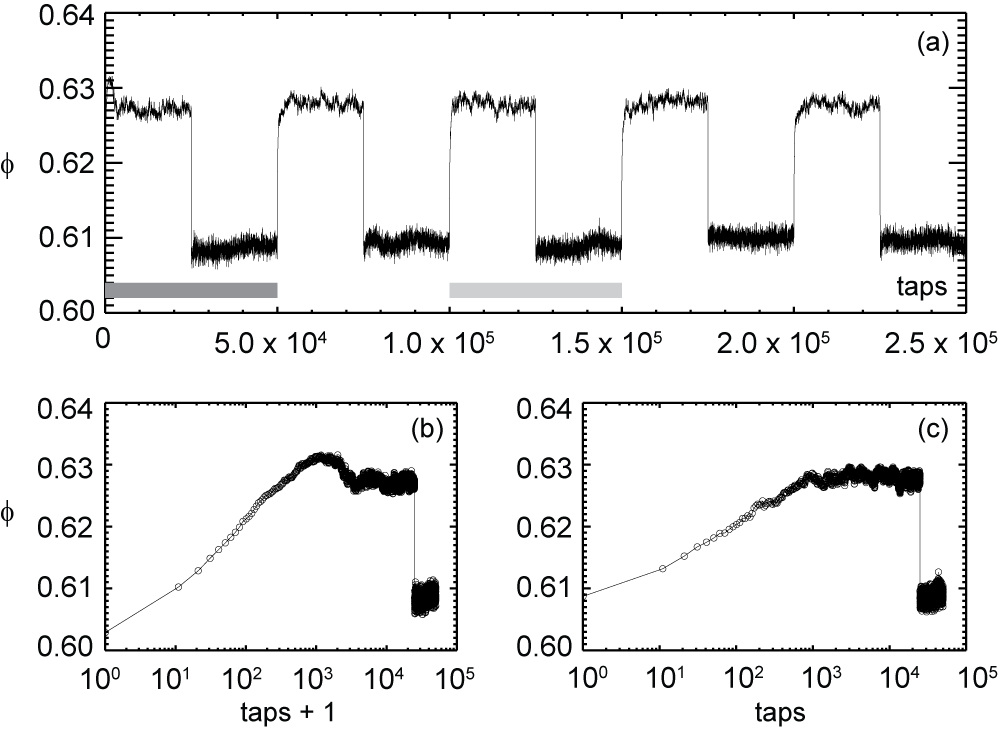}
    \caption{\label{fig:pheno} (a) Evolution of the packing
    fraction as function of the number of taps. Here, we fix
    $T=33$ ms, and $\Gamma$ is alternated between $\Gamma=2$ and
    $\Gamma=6$ every 25000 taps. Our data shows that after a short
    transient, the density can be reversibly changed from one
    value to another.
    (b) The initial transient compaction behavior on a logarithmic scale, for the
    range indicated by the dark grey bar in panel (a). (c) A
    typical compaction process between two steady states as
    occurs at $t=10^5$  as indicated by the light grey
    bar in panel (a).}
\end{figure}

Fig.~\ref{fig:pheno}a shows how the packing typically changes with
the number of taps applied to the system, in an experiment where a
packing of glass beads is subjected to 250000 taps with $T=33$ ms,
and where the amplitude is altered between $\Gamma \sim 2$ and
$\Gamma \sim 6$ after each 25000 taps. The packing fraction is
only measured every 10 taps. There are several distinct features
in this graph: \textit{(i) Initial transient:} Initially the
packing fraction evolves slowly towards a steady state value,
nonmonotonically in the case depicted in Fig.~\ref{fig:pheno}b.
\textit{(ii) Steady state:} After the transient the system is able
to reach a steady state packing fraction. The correspondence
between $\phi$ and $\Gamma$ is also reproducible: changing tap
amplitudes repeatedly back and forth shows that for a particular
tap amplitude the same steady state packing fraction is always
obtained (Fig.~\ref{fig:pheno}a). \textit{(iii) Amplitude step
transient:} When the amplitude of the tap changes, the packing
fraction changes too, but not instantaneously. This time dependent
process is usually not symmetric in the step direction: an
increase of the tap amplitude usually almost immediately dilates
the packing to the packing fraction appropriate for the new
amplitude. A step down in tap amplitude requires compaction of the
packing to reach the steady state packing fraction belonging to
the lower tap amplitude. This process is usually slower. As can be
seen in Fig.~\ref{fig:pheno}c, the increase in density from
roughly 0.61 to the plateau value $\approx 0.63$ takes of the
order of 1000
taps, comparable  to the initial transient duration.\\

The initial transient is somewhat different in each experiment, and
run-to-run fluctuations of details of $\phi(t)$ are considerable.
For example, non-monotonicity seen in $\phi(t)$ in
Fig.~\ref{fig:pheno} is not always observed. This may either be
due to the fact that the preparation of the packing in each
experiment is slightly different, or to inherent fluctuations of
the density evolution. We identify the transient behavior with the
irreversible branch, and the steady state with the reversible
branch.

The strength of the fluctuations in the early evolution in
$\phi(t)$ hinder a precise comparison of our results to the
results from the Chicago and Rennes
groups~\cite{1995_pre_knight,1997_powtech_nowak,2002_epl_philippe}.
However, our values for the densities on the reversible branch,
$\phi_{\rm rev}(\Gamma)$, are far more robust, and in the rest of
the paper we focus on $\phi_{\rm rev}(\Gamma,T)$ in the steady
state.

\begin{figure}[t]
    \includegraphics[width=8cm]{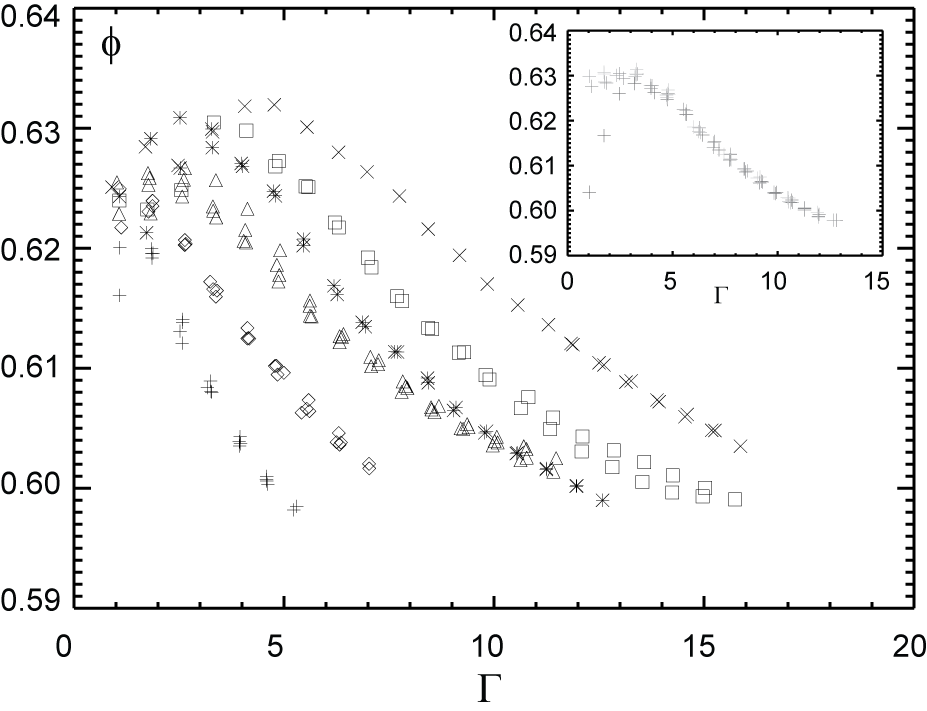}
    \caption{\label{fig:glass} Reversible and irreversible branch
    for the glass bead mixture (See text). The main panel shows
    the reversible branch $\phi_{\rm rev}(\Gamma,T)$ for different tap lengths $T$ ($+$ 50ms, $\diamond$ 33ms, $\triangle$ 16ms, $\ast$ 14ms,
    $\Box$ 13ms, $\times$ 10ms). Clearly $\Gamma$ alone is not
    sufficient to characterize the tapping.
    The inset shows $\phi_{\rm rev}(\Gamma,T)$ for $T$=14.3ms, and four sweeps in $\Gamma$
    (see text). Greyscale of symbols indicates measurement time;
    lighter points are later measurements. After an initial
    transient (black crosses), the densities become history independent.}
\end{figure}

\section{\label{sec:steady}
Steady-State Density as a Function of $\Gamma$ and $T$}

As we will show, our experimental protocol allows the
determination of $\phi_{\rm rev}(\Gamma,T)$. At fixed $T$, this
enables use to reproduce the reversible and irreversible branch in
a $\phi_{\rm rev}(\Gamma,T)$-plot similar to those found before in
the Chicago and Rennes experiments by sweeping $\Gamma$ up and
down several times.

\begin{figure}[t]
    \includegraphics[width=8cm]{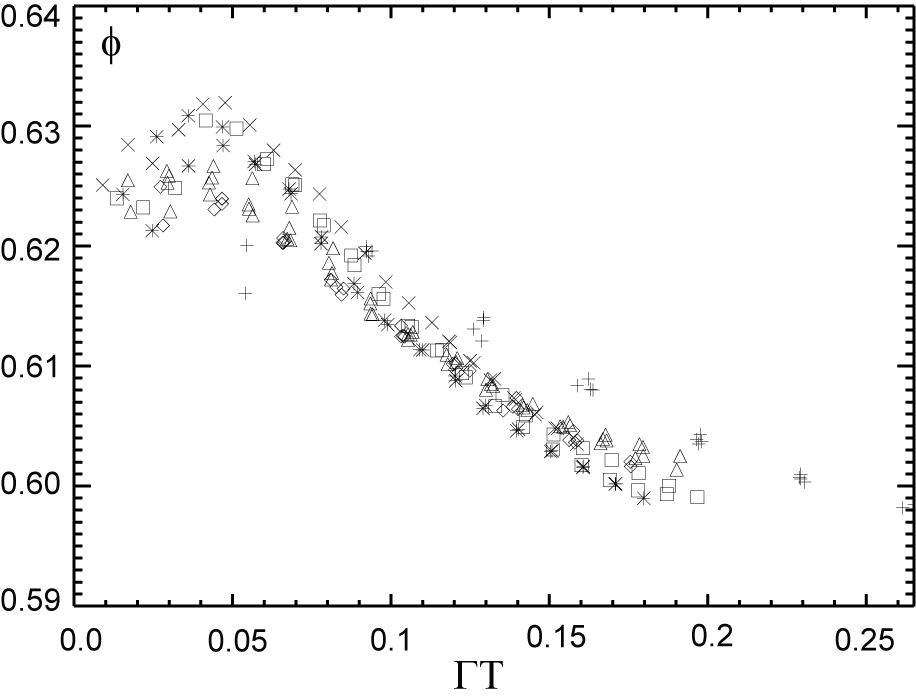}
    \caption{\label{fig:glasscollapse} Data for the reversible
    branches for the glass bead mixture as shown in
    Fig.~\ref{fig:glass} can be collapsed well when plotted as
    function of $\Gamma T$.}
\end{figure}

In the inset of Fig.~\ref{fig:glass} we show that the packing
density becomes well-defined on the reversible branch. In this
experiment, starting from a low density packing, the tap amplitude
is sweeped up and down four times, and the number of taps at each
different tap amplitude is 4000. Every 20 taps the packing
fraction is measured and each data point corresponds to an average
over the resulting 200 measurements. The irreversible branch is
visible as the initial increase of $\phi$ with $\Gamma$. After
about 12,000 taps the reversible branch is reached: this branch is
clearly shown in the four amplitude sweeps that all follow the
same $\phi_{\rm rev}(\Gamma,T)$
relation, with $\phi_{\rm rev}$ decreasing with $\Gamma$.\\

\textit{Effect of the tap duration.} We will now explore the
reversible branch for a range of values of $T$
(Fig.~\ref{fig:glass}).
The reversible branches obtained in  a series of experiments in
which $\Gamma$ is sweeped for a range of values of $T$ are shown
in the main panel of Fig.~\ref{fig:glass}. The reversible branches
for different tap lengths $T$ clearly have the same overall form,
but do not coincide --- $\Gamma$ alone is not the parameter that
governs compaction. The spacing of the data and the fact that the
functional form of $\phi_{\rm rev}(\Gamma,T)$ is similar for
different $T$, strongly suggests the data sets can be collapsed
onto a master curve. Fig.~\ref{fig:glasscollapse} shows that
$\phi_{\rm ref}(\Gamma,T)$ can be collapsed reasonably well by
plotting $\phi_{\rm rev}$ as a function of the product $\Gamma T$.

\begin{figure}
    \includegraphics[width=8cm]{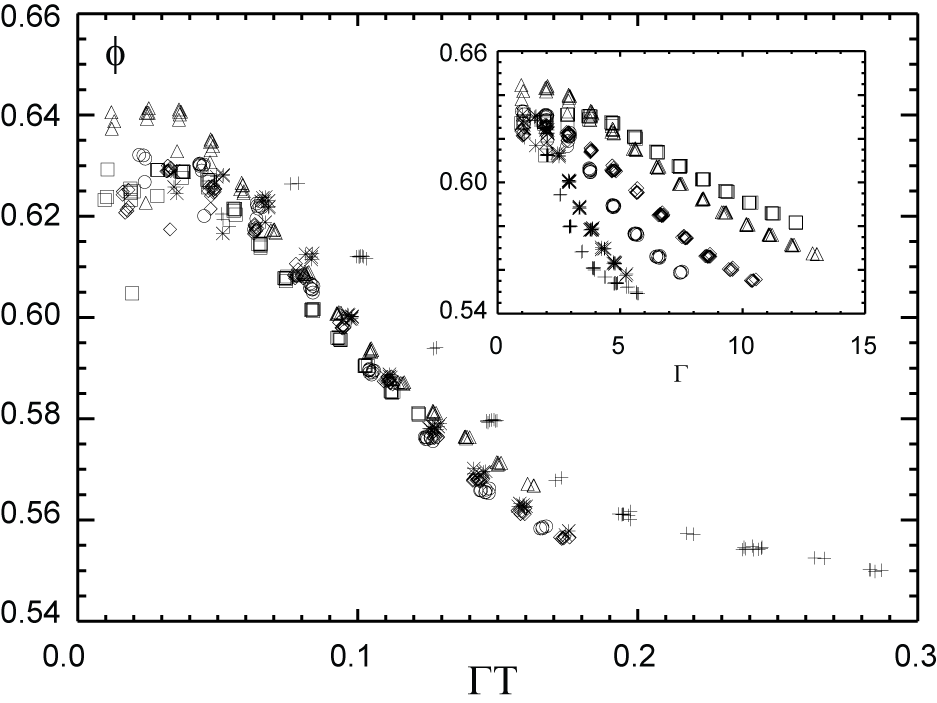}
    \caption{\label{fig:bronze}$\phi_{\rm rev}(\Gamma T)$ for the rough bronze powder. Symbols: $+$ 50ms,
    $\ast$ 33ms, $\circ$ 22ms, $\diamond$ 17ms, $\triangle$ 13ms , $\Box$ 10ms.
    Inset: Same data, $\phi_{\rm rev}(\Gamma,T)$ for the different $T$.}
\end{figure}

\textit{Bronze powder.} For the bronze powder, we measured the
reversible branch $\phi_{\rm rev}(\Gamma,T)$ for different values
of the tap length $T$; the results are shown in the inset of
Fig.~\ref{fig:bronze}. Note that the range of packing fractions
obtained is far larger for the bronze powder, which may be
attributed to the roughness of the particles~\cite{2005_prl_an}.
Similarly to the glass beads, $\phi_{\rm rev}$ collapses onto a
master curve when plotted as function of $\Gamma T$ --- see
Fig.~\ref{fig:bronze}. This shows that the details of the granular
material used are entirely insignificant.

\textit{Absence of hysteresis.} In Ref.~\cite{2000_prl_josserand}
memory effects were observed in the evolution of the packing
fraction on the irreversible branch: two different initial
conditions were prepared at a certain fixed $\phi$, by compacting
low density systems at different tapping strength. The time
evolution of subsequent compaction differed, despite the fact that
the initial packing fraction and tap strength were equal
--- hence not only the packing fraction, but also the history is
important for the evolution at the irreversible branch.

While such memory effects have not been seen, per definition, on
the reversible branch, one may wonder if some more subtle
hysteretic effects could arise there. In particular, if the state
of the system on the reversible branch is not fully specified by
density, it might be that using two control parameters, subtle
hysteretic effects not seen when only sweeping the single
parameter $\Gamma$ become apparent.

Since data for $\phi_{\rm rev}$ as a function of $\Gamma$ and $T$
can be collapsed on a single master curve, it follows that
$\Gamma$ and $T$ can be varied at the same time in such a way that
$\phi_{\rm rev}$ stays constant. As we will show, we do not
observe any appreciable hysteretic effects when switching between
different pairs of ($\Gamma,T$) adjusted so that $\phi_{\rm
rev}(\Gamma,T)$ is constant.

\begin{figure}
    \includegraphics[width=8cm]{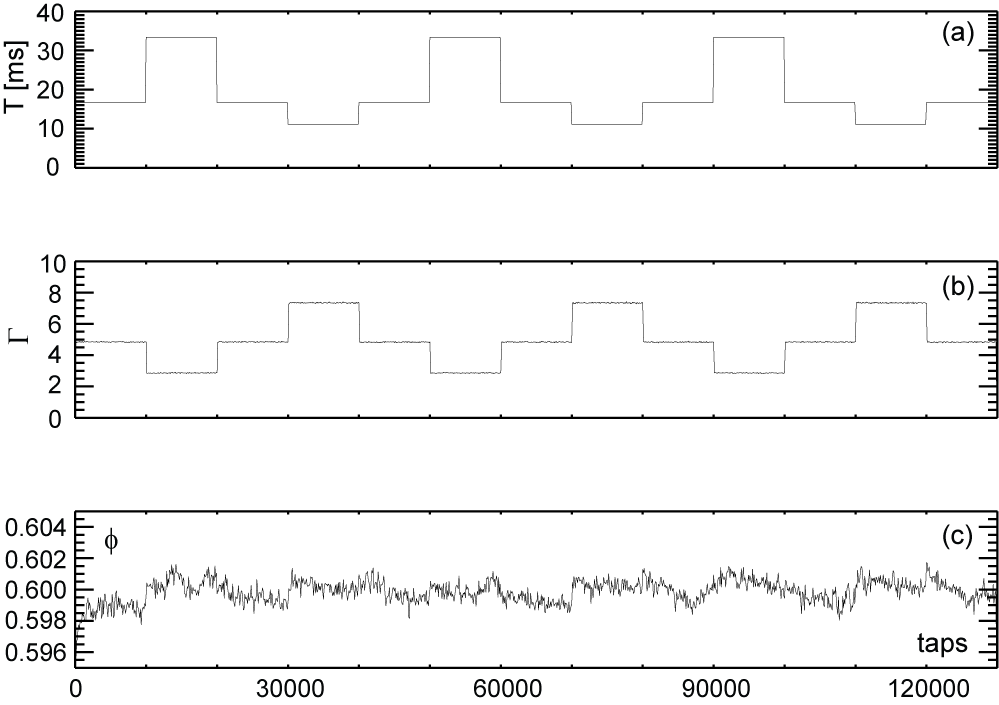}
    \caption{\label{fig:4panelconstdens} $T$ (a) and $\Gamma$ (b) used in
    the constant packing fraction sweep. The packing fraction evolution in time
    is shown in (c); the fluctuations visible in (c) are
    of the order of the experimental noise. For details see text.}
\end{figure}

We measure the packing fraction of a bronze powder packing while
we expose the granular packing to taps with a sequence of
different $T$: 17,33,17,11 ms (Fig.~\ref{fig:4panelconstdens}a)
and different $\Gamma$. This sequence is repeated three times;
each $\Gamma,T$-pair was used for 10000 taps, the total number of
taps applied in the experiment is 130000. The amplitude $\Gamma$
(4.84, 2.88, 4.84, 7.32) for each different $T$, shown in
Fig.~\ref{fig:4panelconstdens}b, was fine tuned such that the
resulting packing fraction stayed the same during the whole
experiment; the product $\Gamma T$ is constant to within 10$\%$.
This 10$\%$ variation suggests that $\Gamma T$ is likely an
effective approximation of the ultimate order parameter that sets
the density on the reversible branch
--- see below.
The packing fraction evolution during the experiment is shown in
Fig.~\ref{fig:4panelconstdens}c. No appreciable transients are
observed when switching $\Gamma$ and $T$: there is no evidence for
any appreciable hysteresis. Our data is further evidence that
memory effects, which are a characteristic of the irreversible
branch, do not play a role on the reversible branch.

\section{\label{sec:interp}Interpretation} How can we
understand the relevance of $\Gamma T$ in determining
$\phi_{rev}$? We note that the liftoff velocity $\dot{z}_l$, the
time-of-flight $\tau_{f}$ and impact velocity $\dot{z}_i$ are all
approximately, although not precisely, proportional to $\Gamma T$.
Below we briefly discuss these parameters, but note already here
that the experimental scatter prevents us from unambiguously
determining whether these provide better scaling collapse than the
simple $\Gamma T$ scaling employed above.

\textit{(i) Liftoff velocity ---} The liftoff velocity $\dot{z}_l$
is determined by calculating the velocity at the time where the
acceleration falls below a threshold $-g\Gamma^*$. In case the
packing experiences no friction with the container walls, this
threshold equals the gravitational acceleration ($\Gamma^*=1$),
but in general $\Gamma^*$ is somewhat larger, typically around
1.2. For given $\Gamma^*$ and waveforms as described above, one
finds that in the parameter regime where liftoff occurs ($\Gamma
> \Gamma^*$), the liftoff velocity is given by
\begin{equation}
\dot{z}_l=\Gamma ~ T \times \left(
\frac{g}{2\pi}\left[1+\sqrt{1-(\Gamma^*/\Gamma)^2} -
\frac{T}{P}\right]\right)~.
\end{equation}
For most parameter values, the term within square brackets is
close to two --- typically, $T/P$ is of order of a few percent,
and for $\Gamma > 5$, the square root term is within a few percent
equal to one. The main deviations between the scaling of $\Gamma
T$ and $\dot{z}_l$ occur for small $\Gamma$. Our scatter is
relatively large in this regime, and in a plot of $\phi_{\rm rev}$
as function of $\dot{z}_l$, the quality of the collapse is very
similar to when $\Gamma T$ is used.

\textit{(ii) Time of flight ---} In rough approximation, the time
of flight $\tau_{f}$ is simply proportional to the liftoff
velocity: $\tau_{f}=2\dot{z}_{l}/g$. We have also calculated the
time of flight numerically, assuming either the analytic
expressions for $z(t)$ and $\ddot{z}(t)$, or using the
experimental data available for $z(t)$ and $\ddot{z}(t)$, and
taking drag forces into account (these affect the total time of
flight of the packing~\cite{2006_pre_lumay}). Despite the presence
of free fit parameters (take-off acceleration, drag coefficient),
the resulting data collapse when $\tau_f$ is used as scaling
parameter is only marginal better than for $\Gamma T$ or
$\dot{z}_l$.

\textit{(iii) Impact velocity ---} The amount of energy dissipated
once the packing comes to rest in the container is set by the
impact velocity, and this could also be an appropriate choice of
rescaling parameter. However, in the range of parameters explored,
the impact velocity scales very similar to the time of flight, and
the quality of the collapse does not improve.

We hence conclude that while impact velocity or time of flight
might give a marginal improvement in the data collapse, the much
simpler rescaling parameter $\Gamma T$ works essentially equally
well.

\section{\label{sec:conc}Conclusions}
By precisely controlling the shape of the taps, we find that the
compaction of vibrated granular media is not only governed by
$\Gamma$, the dimensionless peak acceleration, but that the tap
duration plays an equally important role. We observe a collapse of
the reversible branch with $\Gamma T$, for two different types of
granular materials. We do not see any evidence for hysteretic
effects when switching between different driving parameters that
correspond to the same reversible-branch-density.

The time of flight,  the liftoff velocity and the impact velocity
all scale similar to $\Gamma T$, which makes it impossible to
experimentally determine which parameter leads to the best data
collapse. In experiments focussing on the transients in granular
compaction~\cite{2008_epl_ludewig}, or on the hydrodynamic phases
in vibrofluidize granular materials~\cite{2005_prl_eshuis}, the
order parameter found was always proportional to the injected
energy per vibration cycle. We suggest that our scaling supports
the view that the energy injected is also the driving mechanism in
compaction experiments.


\begin{acknowledgements}
The authors like to thank Jeroen Mesman for technical assistance.
J.A.D. acknowledges support from the Dutch physics foundation FOM,
and M. v. H. acknowledges support from NWO/VIDI.
\end{acknowledgements}



\end{document}